\documentclass[
letterpaper,
showpacs,floatfix, aps,prl,amsmath,amssymb,
twocolumn,
groupaddress
]{revtex4}

\usepackage{graphicx}

\newcommand{\be}{\begin{equation}}
\newcommand{\ee}{\end{equation}}
\newcommand{\bea}{\begin{eqnarray}}
\newcommand{\eea}{\end{eqnarray}}
\newcommand{\bs}{\begin{split}}
\newcommand{\bes}{\begin{equation}\begin{split}}
\newcommand{\ees}{\end{split} \end{equation}}

\newcommand{\D}{\mathcal{D}}
\newcommand{\G}{\mathcal{G}}
\newcommand{\C}{\mathcal{C}}

\newcommand{\N}{\mathcal{N}}
\newcommand{\Hh}{\mathcal{H}}

\renewcommand{\Re}{\mathrm{Re}} 
\newcommand{\Tr}{\mathrm{Tr}}
\newcommand{\gl}{\langle g_1 \rangle}
\newcommand{\gr}{\langle g_2 \rangle}

\newcommand{\req}[1]{Eq.~(\ref{#1})}
\newcommand{\reqs}[1]{Eqs.~(\ref{#1})}

\begin{document}

 \title{Antilocalization in Coulomb Blockade}

\author{Y. Ahmadian and I. L. Aleiner}
\affiliation{Physics Department, Columbia University, New York, NY 10027}


\begin{abstract}
  We study the effect of spin-orbit scattering on the
  statistics of the conductance of a quantum dot for Coulomb
  blockade peaks and valleys. We find the distribution function
  of the peak heights for  strong spin-orbit scattering in the presence and
  absence of time reversal symmetry. We find that the
  application of a magnetic field
  suppresses the average peak height, similar to the antilocalizaion in
  the bulk systems. For the valleys, we consider the elastic cotunneling
  contribution to the conductance and calculate 
its moments
at the crossover between
  ensembles of various symmetries.
\end{abstract}
\date{\today}
\maketitle

Electron transport in quantum dots (QD) is  affected by
electron-electron interactions, the profound example being 
Coulomb blockade (CB) \cite{coulombblockade} in QDs connected
to leads by contacts of small conductance $G\ll e^2/\pi\hbar \equiv
G_q$. Due to charge quantization, at temperatures smaller than the
charging energy $E_c \gg \Delta$ ($\Delta$ being the one particle level
spacing), associated with the addition of one electron to
the QD, transport through the dot is diminished except at
discrete values of the
gate voltage corresponding to  charge degeneracy. 
 Thus, the dependence $G(V_g)$
exhibits almost equidistant sharp peaks separated by deep minima
(CB valleys). Close to the peaks and at low temperatures
$T\ll \Delta$, transport occurs by resonant tunneling through a single
state. The situation is different in the valleys where
transport is due to virtual transitions of electrons via excited
states of the QD: many levels ($N \sim E_c/\Delta$) contribute to
tunneling. In both
cases, the mesoscopic fluctuations of the conductance are determined
by statistics of wavefunctions in the closed dot.

In disordered or chaotic QDs 
with the Thouless energy $E_{\mathrm{Th}}$ 
much larger than all relevant energy
scales, transport phenomena are described by Random Matrix Theory
(RMT) (see e.g. \cite{aleiner:brouwer:glazman:02}).  In this theory,
the statistics of one particle energy levels and eigenfunctions are
described by universal ensembles  sensitive only to the
underlying symmetries. 
The statistical distribution of the peak heights has been calculated
 based on RMT for
the Gaussian orthogonal (GOE) and unitary (GUE)
ensembles \cite{jalabert:stone}. The results were extended to the crossover between
those ensembles \cite{alhassid:etal}, using the statistics of
wavefunctions derived in Ref.
\onlinecite{falko:efetov:prl96}.  The average peak height was observed
to increase with the magnetic field, reminiscent of
weak localization in  bulk systems \cite{chang:etal:96}. The statistics of conductance in
the valleys was studied theoretically \cite{aleiner:glazman:96} and
experimentally \cite{cronenwett:etal:97} for the same crossover.
However, there is a gap in the theoretical literature regarding the
effect of spin-orbit scattering on these statistics. The effect of
breaking of time reversal symmetry on the CB peak
heights in quantum dots with strong spin-orbit scattering is expected
to be strong, similar to
antilocalization in bulk systems \cite{bergmann:84} 
or open dots \cite{zumbuhl:so:01}.

In this Letter, we find the statistics of the peak heights for strong
spin-orbit scattering in the presence and absence of time reversal
symmetry. For the valleys we calculate the magnetic correlation for conductance, 
even in the crossover  between ensembles of various
symmetries.

The QD attached to two leads, $1$ and $2$, 
is described by the Hamiltonian
$
H = H_L   + H_D + H_{LD},
$
where 
\be
H_L = v_F \sum_{a = 1,2} \sum_{\sigma}\int\frac{dk}{2\pi} k c^\dagger_{a,\sigma}(k) c_{a, \sigma}(k),
\ee
corresponds to the leads, with $v_F$ being the Fermi velocity, 
and $\sigma=\uparrow,\downarrow$ labels the spin. 
We  consider only single channel 
contacts. 
The dot Hamiltonian is  
\be
H_D = \sum_{m, n=1}^{M} \sum_{\sigma_1\sigma_2}
c^\dagger_{m,\sigma_1}
 \Hh_{mn}^{\sigma_1\sigma_2} c_{n,
\sigma_2} + E_c (\hat{n} - \N)^2, 
\ee
where the first term describes the non-interacting dynamics of the
closed dot, 
and the second term, with
$\hat{n} = \sum_{m,\sigma} c^\dagger_{m,\sigma} c_{m,\sigma}$,
corresponds to charging energy.
Dimensionless parameter $\N$ is a linear function of the gate voltage. 
The matrix $\Hh$ belongs to an RMT ensemble. 
In the case of quantum dots based on a 2D electron gas,
 the spin-orbit interaction is atypical and is
characterized by two parameters $\epsilon_\perp^{\mathrm{so}}$ and
$\epsilon_\parallel^{\mathrm{so}}$ \cite{aleiner:falko}. In the
presence of a magnetic field with both perpendicular and parallel
components to the gas plane (let the third axis be normal to the
plane),  $\Hh$ can be written as \cite{cremers:brouwer:falko}
\bea
\label{rmthamil}
\Hh =  \frac{\Delta}{2\pi} &\big[& 
 \Hh_0 \mathbf{1} + i\mathcal{X} 
(x\mathbf{1} + a_\perp \sigma^3) + 
i a (\mathcal{A}_1\sigma^1 + \mathcal{A}_2\sigma^2) \nonumber\\ 
& 
+& b_\perp \mathcal{B}_h \sigma^3\big] - 
\epsilon_Z \vec{l} \cdot \vec{\sigma}/2,
\eea
where $\Hh_0$ and $\mathcal{B}_h$ are real symmetric $M\times M$ matrices, with
$\langle\Tr \Hh_0^2\rangle = M^3$, $\langle\Tr \mathcal{B}_h^2\rangle
= M^2$, and $\mathcal{X}$ and $\mathcal{A}_i$ are real antisymmetric
matrices with $\langle\Tr \mathcal{X}\mathcal{X}^T\rangle = M^2$ and
$\langle\Tr\mathcal{A}_i \mathcal{A}_j^T \rangle = M^2\delta_{ij}$ (the limit $M\to \infty$ is taken eventually).
Here $a_\perp^2 = \pi \epsilon_\perp^{\mathrm{so}}/\Delta$, $a^2 = \pi
\epsilon_\parallel^{\mathrm{so}}/\Delta$, $\epsilon^Z$ is the Zeeman
splitting energy due to the parallel magnetic field in the direction
given by the unit vector $\vec{l} = (l_1,l_2,0)$, and $b_\perp^2 =
\pi\epsilon_\perp^Z/\Delta$ where $\epsilon_\perp^Z$ describes the
combined effect of Zeeman splitting and spin-orbit scattering. 
The orbital effect of the magnetic field is
characterized by the energy  $\epsilon_B = x^2 \Delta/\pi =
\kappa E_{\mathrm{Th}}(\Phi/\Phi_0)^2$, where $\kappa$ is a
coefficient  dependent on the shape of QD, 
$\Phi$ is the flux of the magnetic field through
the dot, and $\Phi_0$ is the flux quantum.

The tunneling Hamiltonian $H_{LD}$ couples the 
states of the  leads $a=1,2$ to orbital states in the QD: 
\bea\label{coupling}
H_{LD} =\sqrt{\frac{M \Delta}{\pi^2 \nu}} \sum_{a, n,k} 
 \delta_{a n}
t_n c^\dagger_a (k) c_n + h.c.,
\eea
Here $\nu=1/(2\pi v_F)$ is  the leads density of states  per spin.  

First we consider the statistics of the peaks. For the temperature range $T\ll \Delta \ll E_c$, 
only the last occupied, possibly degenerate level participates 
in the electron transport close to the peaks. 
Using the Golden Rule,  the escape rates of the 
level $\alpha$ into the first and second leads are $\Gamma^\alpha_{1,2} 
=\frac{\Delta}{2\pi\hbar} g_{1,2}$, where $g_{i} = 4 |t_{i}|^2 y_i$ is
the dimensionless conductance of the $i$-th contact. Here
\be\label{yi} y_i = \frac{g_i}{\langle g_i \rangle} = M
\psi^\dagger_{\alpha}(i)\psi_\alpha(i), \qquad (i = 1, 2), 
\ee is a
fluctuating quantity, and $\psi_\alpha(i)$ are spinors with components $ \langle i,\uparrow \downarrow |
\alpha \rangle$. For the peaks we will confine ourselves to strong spin-orbit scattering
$\epsilon_\parallel^{\mathrm{so}}, \epsilon_\parallel^{\mathrm{so}}
\gg \Delta$. In this case, and for zero magnetic field, the random
matrix $\Hh$ belongs to a Gaussian symplectic ensemble (GSE): it is a
Hermitian matrix with quaternionic entries. In the notation of
\cite{aleiner:falko} this corresponds to the symmetry class $(\beta =
4, \Sigma = 1, s =2)$. The number $s=2$ signifies Kramer's degeneracy
of energy levels. In the presence of a
strong enough magnetic field such that either $\epsilon_B\gg \Delta$ or
$\epsilon_Z\gg \Delta$, the
Hamiltonian $\Hh$ belongs to a GUE corresponding
to $(\beta=2,\Sigma=2, s=1)$, and the symmetry group of the RMT
ensemble is $\mathrm{U}(2M)$. In this case, the levels are not
Kramer's degenerate.  In both cases, the random variables $y_i$ in
\req{yi} are statistically independent, and their distribution is
given by \be\label{proby} p(y) = 4 y e^{-2 y}.  \ee

Assuming 
$\Delta\gg T \gg \Gamma^\alpha = (g_1 + g_2)\Delta/2\pi\hbar$,
one can use the rate equations
\cite{aleiner:brouwer:glazman:02,rateequations}.  
In the
GSE case, we will label the Kramer's degenerate levels by
$\psi_{\alpha, s}$, with the $s=\pm$ states related by the operation
of time reversal: $\psi_{\alpha -} = i\sigma^y \psi_{\alpha +}^*$. Due
to this relation, $(\psi_{\alpha,+}^\dagger\psi_{\alpha,+}) =
(\psi_{\alpha,-}^\dagger\psi_{\alpha,-})$, and the escape rates are
therefore equal for the $s=\pm$ states (see
\req{yi}).  In the absence of interactions, the level $\alpha$ can be
in four states; empty, doubly occupied, or singly occupied with $s=\pm$. The picture
changes when we take the interaction into account: due to the large
charging energy, only three states with the number of electrons on the
dot differing by $1$ can be in resonance. For the peaks
corresponding to $\N^* = 2j + 1/2$,  ($j\in \mathbb{Z}$), the
doubly occupied state has an extra energy $E_c$, and does not
participate in transport (the case $\N^* = 2j - 1/2$ gives the same
result for the peak height). In the
stationary state  the rate equations yield
\be\label{gofn}
G({\cal N}) = \frac{I}{V}\Big|_{V\to 0} = -  G_q \frac{\Delta}{T} \frac{g_1 g_2}{g_1 + g_2} \frac{\partial f_F/\partial x}{1+f_F(x)},
\ee
where $f_F(x)  = 1/(1+ \exp{x})$, and $x= 2E_c (\N - \N^*)/T$. 
At the maximum which slightly deviates from $x=0$
\be
G^{\beta=4,s=2}_{peak}/G_q = (3-2^{3/2})   \frac{\Delta}{T} \frac{g_1 g_2}{g_1 + g_2}.
\ee

\begin{figure}[!bt]
\includegraphics[height=1.65in,width=2.4in,angle=0]{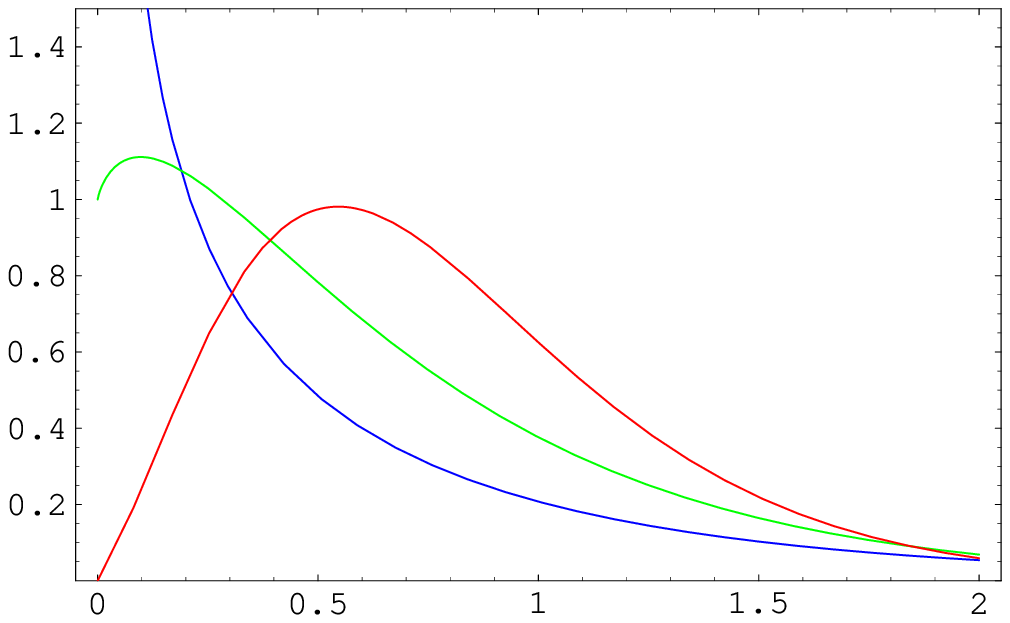}
\put(-5,-5){$\alpha$}
\put(-199,110){$W(\alpha)$}
\put(-145,105){$\beta=1$}
\put(-115,85){$\beta=4$}
\put(-120,55){$\beta=2$}
\caption{The probability 
distribution $W(\alpha)$ in different ensembles, 
for symmetric contacts ($a=0$). The $\beta=1,2$, 
were studied in Refs. \onlinecite{jalabert:stone}.}
\label{fig1}
\end{figure}

The only difference in the $(\beta=2,\Sigma=2, s=1)$ case is that now
there is no Kramers' degeneracy and the resonant level can only be in
two states: empty or occupied. 
Similarly to \req{gofn} we find 
\be\label{gofn2}
G(\N) = - \frac{G_q}{2} \frac{\Delta}{T} \frac{g_1 g_2}{g_1 + g_2} \frac{\partial f_F(x)}{\partial x}.
\ee 
The maximum occurs at $x=0$, and we find
\be
G^{\beta=2,s=1}_{peak}/G_q= \frac{1}{8}  \frac{\Delta}{T}\frac{g_1 g_2}{g_1 + g_2}.
\ee
Since the $y_i$'s have the same distribution in both ensembles we see that  $G_{peak}$ has the same distribution in both cases except for a scaling. 
Using the probability distribution \req{proby}, we obtain $G_{peak}$ in terms of the average contudctances $\langle g_{1,2}\rangle$ and a single random variable $\alpha$, 
\be\label{resultofg}
G^{\beta,s}_{peak}/G_q =
 \alpha  \frac{\Delta}{T}\frac{2\gl\gr}{(\gl^{1/2} + \gr^{1/2})^2} \chi_{\beta,s},
\ee
where $\chi_{\beta=4,s=2} = 3-2^{3/2}$, and $\chi_{\beta=2,s=1} = 1/8$. 
The probability distribution for $\alpha$ is given by (see Fig. \ref{fig1})
\bea
W(\alpha) &= &  
16 \alpha^3 (1-a)^4 e^{-2\alpha(1+a)} \bigg\{ \frac{1+b^2}{2} K_0
[2\alpha(1-a)] 
\nonumber\\
&+& \left[b + \frac{b^2-1/2}{2\alpha(1-a)}\right] K_1
[2\alpha(1-a)]\bigg\}.
\label{probalp}
\eea
Here $K_{0}(x)$ and $K_1(x)$ are 
MacDonald functions 
and 
\bea
a = \left(\frac{\gl^{1/2} - \gr^{1/2}}{\gl^{1/2} +
    \gr^{1/2}}\right)^2,
 &\qquad& b=\frac{1+a}{1-a}.
\eea
For the average we obtain
\bea
\label{average}
\langle \alpha \rangle = 8 (1-a)^4
\left\{ \frac{2(1+b^2)}{105}F(5,1/2,11/2;a) \right. \qquad\qquad && 
\\ \left. + \frac{1+a}{21}F(6,3/2,11/2;a) + \frac{b^2-1/2}{35} 
F(5,3/2,9/2;a)\right\},&&\nonumber
\eea
with $F(\dots)$ being the hypergeometric functions.
Equation (\ref{average}) is well approximated by $\langle \alpha \rangle \simeq (8-3a)/10$.

Equation (\ref{resultofg})
is our main result for the statisitical distribution of
the peak heights. We see that the application of the
magnetic field  causes the average conductance to drop by a
factor $8(3-2^{3/2})\simeq 1.37$, similar to antilocalization for bulk
systems. Surprisingly, the shape of the distribution remains the same. This
is because at strong magnetic field the RMT ensemble
crosses over to a unitary ensemble with two channel leads corresponding to
spin projections. The statistics of the eigenstates in the latter are
the same as that of the symplectic ensemble, and hence the same
distribution of peak heights.  Though the drop in the average peak heights is
a manifestation of the lifting of the Kramer's degenaracy, the
numerical
factor for this drop is non-trivial.
Had we ignored the charging energy, we would instead
obtain a reduction by a factor of $2$. 
The different numerical factor originates in the exclusion of 
the doubly occupied degenerate
level from transport by the electron-electron interaction. We emphasize, however,
 that the distribution
function changes for intermediate values of the
perpendicular field. 


Next we turn to the mesoscopic fluctuations of the valleys, 
$|\N-\N^*| > T/E_c$. Then, \reqs{gofn}-(\ref{gofn2}) suggest an
exponentially small conductance.  However, this is incorrect,
as the rate equations are based on processes  of the first order
in $H_{LD}$. For such processes a
charge is transfered between a lead and QD,  and the conductance
in the valleys is small because of the charging gap in the final state. 
This is not the case for higher order processes
that allow for both the initial and the final excitations to be in the
leads, so there is no gap for the final state. Such processes are
referred to as co-tunneling
\cite{aleiner:brouwer:glazman:02,cotunneling}, 
and are suppressed
only algebraically, $G(\N) \propto 1/(2E_c|\N - \N^*|)$. We  consider
elastic co-tunneling which is the dominant process in the temperature
range $T \ll (\Delta E_c)^{1/2}$. In these processes a charge is transferred from
lead to lead via a virtual transition to an excited state in the dot.
The conductance calculated using the Golden rule is
\cite{aleiner:glazman:96, aleiner:brouwer:glazman:02} 
\be\label{gval}
G= 2\pi^2 G_q \nu^2 \sum_{\sigma_1, \sigma_2 = \pm 1/2}|A_e^{\sigma_1
  \sigma_2} + A_h^{\sigma_1 \sigma_2}|^2, 
\ee 
where the amplitudes
$A_e^{\sigma_1 \sigma_2}$ ($A_h^{\sigma_1 \sigma_2}$) correspond to
processes in which an electron(hole) in spin state $\sigma_1$ in the
first lead is transferred to the second lead in spin state $\sigma_2$.
The second order perturbation theory in
$H_{LD}$ gives 
\bes 
\frac{A_{e(h)}^{\sigma_1 \sigma_2}}{\Delta M } =
\frac{\left(\gl\gr\right)^{1/2}}{4\pi^2\nu}  \sum_{\alpha, \sigma_i}
\frac{\psi_\alpha^{\sigma_2}(2)\psi_\alpha^{ \sigma_1
    *}(1)}{\varepsilon_\alpha \pm E_{e(h)}}
\theta(\pm\varepsilon_\alpha),
\end{split}
\ee
where $\alpha$ is summed over energy levels of the QD.
Here 
\be \label{ec}
E_e = 2 E_c (\N^* - \N),\
E_h =  2E_c (\N - \N^* +1),
\ee
($\N^*-1<\N<\N^*$), are the electrostatic part of the energy of the virtual state for the electron-like and hole-like processes respectively. The eigenenergies $\varepsilon_\alpha$ are measured from the last occupied level so that the step functions $\theta(\pm\varepsilon_\alpha)$ select empty(filled) states for electron(hole) like processes. Equation (\ref{gval}) can be expressed in terms of the Green functions (GF) for the closed dot:
\bea\label{valleyg}
G = G_q \frac{\gl\gr}{4\pi^2}\frac{1}{2}\sum_{\sigma_1, \sigma_2 = \pm 1/2} |F_e^{\sigma_1 \sigma_2} + F_h^{\sigma_1 \sigma_2}|^2,\\
\label{fing}
\frac{F_{e(h)}^{\sigma_1 \sigma_2}}
{M\Delta}
 =  \int_{-\infty}^\infty \frac{d\epsilon}{2\pi i} \frac{ \G^A_{12,\sigma_1 \sigma_2}(\epsilon) - \G^R_{12,\sigma_1 \sigma_2}(\epsilon)}{\epsilon \pm E_{e(h)}}\theta(\pm \epsilon).
\eea
Using this relation we can express all the moments of the conductance
in terms of the average of a product of GFs.  Furthermore
the condition $\Delta \ll E_e,E_h$, allows us to use the diagrammatic
technique in terms of diffusons and cooperons to calculate the
latter. In this approximation the GFs become Gaussain
variables
 with average $\langle \hat{\G}^{R(A)}_{nm}\rangle \propto
 \delta_{nm}$, 
and variances given in terms of the Diffuson and Cooperon matrices
\cite{aleiner:brouwer:glazman:02,aleiner:glazman:96}
{\setlength{\arraycolsep}{-5pt}
\bea
\label{diffuson}
&&\Tr\left\langle \sigma^{\mu} 
\G^R_{12}(\epsilon_1;B_1)\sigma^{\nu}\G^A_{21}(\epsilon_2;B_2)\right\rangle =
\frac{4\pi}{M\Delta^2}\mathcal{D}_{\mu\nu}^\omega(B_{1,2})
\nonumber\\
&&\Tr\left\langle \sigma^{\mu} \tilde{\G}^R_{12}(\epsilon_1;B_1)
\sigma^{\nu}\G^A_{12}(\epsilon_2;B_2)\right\rangle =
\frac{4\pi}{M\Delta^2}\mathcal{C}_{\mu\nu}^\omega(B_{1,2}),
\eea
where  $\omega = \epsilon_1 - \epsilon_2$, $\sigma^{\mu=0}=\hat{1}$, and $\sigma^i$, $i=1,2,3$ are Pauli matrices. Here we defined $\tilde{\G}^R_{12}  = \sigma^2\left(\G^R_{12}\right)^T\sigma^2$. For the Hamiltonian \req{rmthamil}, the inverse of the diffuson and cooperon matrices are given by \cite{aleiner:falko,cremers:brouwer:falko}
\bea
&&\mathcal{D}^{-1}(\omega;B_1,B_2) =   -i\omega\hat{\mathbf{1}} + 
i\epsilon^Z \vec{l}\cdot\vec{\hat{S}} 
\nonumber\\ 
&& \quad + \left(
  \sqrt{\epsilon_B^D}\hat{\mathbf{1}} +
  \sqrt{\epsilon_\perp^{\mathrm{so}}}\hat{S}_3\right)^2
+\epsilon^{\mathrm{so}}_\parallel \left( \hat{S}_1^2 +
  \hat{S}_2^2\right)
 + \hat{S}_3^2 \epsilon_\perp^Z,
\nonumber
\\
&&\mathcal{C}^{-1}(\omega;B_1,B_2) =   -i\omega\hat{\mathbf{1}} +
i\epsilon^Z \hat{\eta}  + \left( \sqrt{\epsilon_B^C}\hat{\mathbf{1}} + 
\sqrt{\epsilon_\perp^{\mathrm{so}}}\hat{S}_3\right)^2 
\nonumber
\\ 
&&
\quad
+\epsilon^{\mathrm{so}}_\parallel \left( \hat{S}_1^2 +
  \hat{S}_2^2\right) 
+ \left(\hat{\mathbf{1}}  -\hat{S}_3^2\right) \epsilon_\perp^Z,
\label{diffcoopinv}
\eea
where $\hat{S}_i^{jk} = - i\varepsilon^{ijk}$, and $\hat{\eta}_{\mu\nu} = l_\mu \delta_{\nu,0} + l_\nu\delta_{\mu,0}$, with $\vec{l}$ the unit vector in the direction of parallel field, and 
\[
\epsilon_B^D \!= \!
\kappa
E_{\mathrm{Th}}\!\left(\frac{B^\perp_1-B^\perp_2}{2\Phi_0/\mathcal{A}}\right)^2, 
\ \epsilon_B^C \!=\! 
\kappa E_{\mathrm{Th}}\!\left(\frac{B^\perp_1+B^\perp_2}{2\Phi_0/\mathcal{A}}\right)^2,
\] 
where $\mathcal{A}$ is the dot's area.
}

\begin{figure}[!bt]
\includegraphics[height=1.65in,width=2.4in,angle=0]{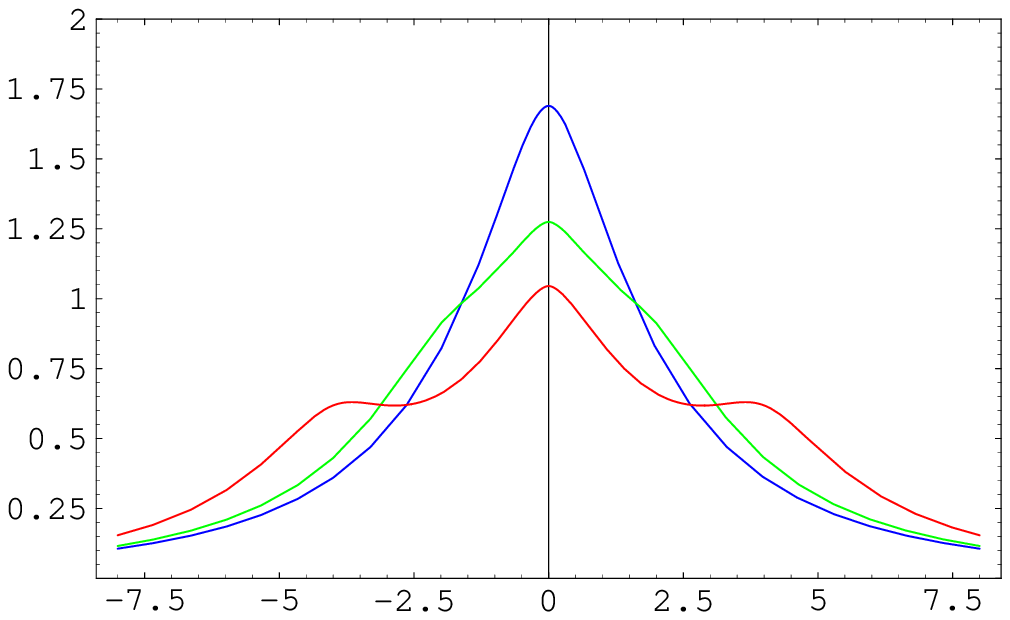}
\put(-100,-8){$\Delta B/B_c$}
\put(-210,110){$C_{12}(B,0)$}
\put(-96,95){$a_\perp=0$}
\put(-96,72){$a_\perp=1$}
\put(-96,57){$a_\perp=2$}
\caption{The correlation function $C_{12}$ for $B_\parallel=0$, $\epsilon_\parallel^{\mathrm{so}}/E = 0.1$, and $a_\perp = \sqrt{\epsilon^{\mathrm{so}}_\perp/E} = 0,1,2$.  The function is non-analytic at $\Delta B/B_c = 0, \pm 2a$, where $B_c \mathcal{A} =\Phi_0 \sqrt{E/\kappa E_{\mathrm{Th}}}$. }
\label{fig2}
\end{figure}

 According to \req{fing} the amplitudes $F_{e}, F_h$ are linear in $\G^R$ and $\G^A$ and are also Gaussian variables with zero average, so that we can calculate the moments of the conductance, \req{valleyg}, using Wick's theorem. From \reqs{valleyg} and (\ref{diffuson}) for $\mu=\nu=0$, we obtain 
 the result
 \be
 \langle G\rangle = 
G_q \frac{\gl\gr}{4\pi^2}\left[\frac{\Delta}{E_e} + \frac{\Delta}{E_h}\right],
 \ee
 independent of all the crossover parameters. 
  Using \reqs{diffuson}-(\ref{diffcoopinv}) together with
  (\ref{valleyg}), we calculate the correlation of the conductances at
  different values of the perpendicular magnetic field. In the
  vicinity of the peaks corresponding to the condition $E_h \ll E_e$
  or $E_e\ll E_h$, for $C_{12}=\langle \delta G(B_1) \delta
  G(B_2)
\rangle/\langle G\rangle^2$ we obtain
\be
C_{12}=\!\!\!\sum_{\substack{\alpha,\beta=1,\dots,4\\i=C,D}} 
\!\!\!\Lambda\left(\frac{\lambda^i_\alpha}{E}\right)
\Lambda\left(\frac{\lambda^i_\beta}{E}\right) 
\frac{\left[\delta_{\alpha\beta} + A^i_{\beta\alpha}A^i_{\alpha\beta}\right]}{8}.
\ee
where $\Lambda(z)=\tilde{\Lambda}(z)-\tilde{\Lambda}(-z)$ and
\be
\tilde{\Lambda}(z)  = \frac{1}{ \pi z} \left[ \ln{(iz)} \ln{(1-iz)} 
  +\frac{1}{4}  \mathrm{Li}_2(-z^2)\right],
\ee
for complex $z$, with $\mathrm{Li}_2(x)$ being the dilogarithm function. The asymptotic behavior is $\Lambda(z) = 1+ z\ln{z}/\pi$, for $|z|\ll 1$, and $\Lambda(z) = (\pi z)^{-1}\ln{}^2 z$, for $\Re\,z\gg 1$. Here $E=\mathrm{min}(E_e,E_h)$, and $\lambda^D_\alpha$ and $\lambda^C_\alpha$ are the four eigenvalues of $\D^{-1}(0;B_1,B_2)$ and $\C^{-1}(0;B_1,B_2)$ matrices respectively. The $4\times 4$ matrices $A_{D,C}$  are given by
\bea
A_D = U_D^{-1}L_D U_D, && A_C = U_C^{-1}L_C U_C,
\eea
where $U_D$ and $U_C$ are the matrices whose columns are the eigenvectors of $\D^{-1}(0;B_1,B_2)$ and $\C^{-1}(0;B_1,B_2)$ respectively, $L_D^{\mu\nu} = \delta^{\mu\nu} - 2\delta^{3\mu}\delta^{3\nu}$ and $L_C^{\mu\nu} = \delta^{\mu\nu} - 2\delta^{0\mu}\delta^{0\nu}$. For zero parallel field and small perpendicular fields Zeeman splitting can be ignored and the eigenvalues of inverse cooperon and diffuson correspond to the usual singlet and triplet states, and we have (see Fig. \ref{fig2})
\bes
C_{12}= &\frac{1}{4}\sum_{i=D,C}\sum_{j=0,1}\sum_{m=-j}^j \left[\Lambda\left(\frac{\lambda^i_{jm}}{E}\right)\right]^2 
\\
\lambda^{D(C)}_{jm} =& \left( \sqrt{\epsilon_B^{D(C)}}+ m\sqrt{\epsilon_\perp^{\mathrm{so}}}\right)^2 \!+\epsilon^{\mathrm{so}}_\parallel \left( j(j+1) - m^2\right). \nonumber
\end{split}\ee

Furthermore for strong spin-orbit scattering, $\epsilon^{\mathrm{so}}_\perp, \epsilon^{\mathrm{so}}_\parallel \gg E$, we can calculate all the moments of the conductance $G$, and obtain the distribution functions for $\gamma = G/\langle G \rangle$:
\be
P(\gamma) = \theta(\gamma) \frac{e^{-4 \tilde{\gamma}}}{\lambda^2} \bigg[  8 \gamma \cosh{4\lambda \tilde{\gamma}}  - \frac{1-\lambda^2}{\lambda/2} \sinh{4\lambda \tilde{\gamma}} \bigg],
\ee
where $\lambda = \Lambda\left(\Phi^2/\Phi_c^2\right)$, $\tilde{\gamma} = \gamma/(1-\lambda^2)$, 
$\Phi$ is the magnetic flux, and $\Phi_c = \Phi_0\sqrt{E/\kappa E_{\mathrm{Th}}} $.

In conclusion, we studied the statistics of the conductance for CB peaks and valleys in the presence of spin-orbit scattering. We calculated the distribution function of the peak heights for strong spin-orbit scattering in the presence and absence of time-reversal symmetry. We found that the average peak height is reduced in the latter case. For the valleys we calculated the average conductance and the correlation function of conductance for a 2DEG QD, as a function of perpendicular magnetic field. 

We are grateful to C. M. Marcus for suggesting this problem.

\end{document}